\documentstyle[aps,multicol,amsmath,epsfig]{revtex}

\begin{document}

\date{May 11, 2000}
\draft

\title{Few-Particle Effects in Semiconductor Quantum Dots:
Observation of Multi-Charged-Excitons}

\author{Arno Hartmann,$^1$ Yann Ducommun,$^1$ Eli Kapon,$^1$
        Ulrich Hohenester,$^2$ and Elisa Molinari$^2$}

\address{
$^1$ Department of Physics, 
     Swiss Federal Institute of Technology Lausanne (EPFL), 
     Switzerland\\
$^2$ Istituto Nazionale per la Fisica della Materia (INFM) and
     Dipartimento di Fisica\\
     Universit\`a di Modena e Reggio Emilia, 
     Via Campi 213/A, 41100 Modena, Italy}

\maketitle

\begin{abstract} 

We investigate experimentally and theoretically few-particle effects in
the optical spectra of single quantum dots (QDs). Photo-depletion of
the QD together with the slow hopping transport of impurity-bound
electrons back to the QD are employed to efficiently control the number
of electrons present in the QD. By investigating structurally identical
QDs, we show that the spectral evolutions observed can be attributed to
intrinsic, multi-particle-related effects, as opposed to extrinsic
QD-impurity environment-related interactions. From our theoretical
calculations we identify the distinct transitions related to excitons
and excitons charged with up to five additional electrons, as well as
neutral and charged biexcitons.

\end{abstract}


\begin{multicols}{2}
\narrowtext

Quantum confinement in low-dimensional semiconductors has been shown to
profoundly affect Coulomb correlations among charge carriers. In
two-dimensional (2D) quantum wells (QWs), enhanced electron-hole
correlations yield stable excitons that dominate the optical absorption
and emission spectra near the band edge \cite{1}. In 1D quantum wires
(QWRs), excitons play an even more important role due to the reduced
Sommerfeld factor \cite{2}, and their dominance in optical spectra was
observed across many interband transitions \cite{3}. In quantum dot
(QD) systems, the importance of Coulomb correlations varies
considerably as a function of the dot size $L$ due to the difference
between the $1/L$ dependence of the Coulomb potential versus the $\sim
1/L^2$ dependence of the confinement energy. Many-particle states can
dramatically change the electronic spectra of QDs compared to the
simple-minded single particle picture of these fully confined states.

Experimentally, the role of Coulomb correlation and many-body effects
in quantum nanostructures has been extensively studied using different
techniques.  Evidence for formation of few-electron states in QDs was
provided by capacitance and by far-infrared spectroscopies \cite{5}.
Multi-exciton states were observed in the luminescence spectra of QDs
formed in QWs and QWRs due to interface disorder \cite{6} and of QDs
produced by Stranski-Krastanow island growth \cite{7}. The formation of
charged excitons in doped QWs was also reported \cite{8}. In the
present Letter, we report the observation of multi-charged exciton
states in the photoluminescence (PL) spectra of QDs with controlled
structure and composition. The binding energies and PL-fine-structure
of the multi-particle states incorporating up to six electrons are
found to be in good agreement with a theoretical model.

Our QDs are fabricated by epitaxial growth on (111)B-oriented GaAs
substrates patterned with an array of micron-sized tetrahedral recesses
\cite{9,10}. Deposition of AlGaAs/GaAs/AlGaAs QW-layers results in the
self-formation of a GaAs QD exactly at the sharp tip of each
tetrahedral recess. Thus, each QD's position is precisely controlled by
the placement of the recess-patterns while its size is controlled by
the growth parameters \cite{11}. Based on atomic force microscopy
studies \cite{12}, we estimate the thickness and diameter of the
approximately lens-shaped QDs discussed here to be 5 and 20 nm,
respectively, which gives rise to an energy splitting of 45 meV between
ground- and excited-state interband transitions. After growth, we
remove the GaAs- substrate from our samples and obtain upright-standing
pyramids with the QD embedded within the pyramid-tip (Fig. 1a). For
luminescence measurements, we use a standard microscopic PL ($\mu$PL)
setup with a $\le 1\mu$m-diameter Ar$^+$-ion laser spot, which can be
easily centered on a single pyramidal structure, thus permitting the
selection of a single QD.

In the present study, the QDs are modulation-doped due to a uniform
background doping of the AlGaAs barrier (x$_{\rm Al}$=0.45) obtained by
organometallic chemical vapor deposition at the growth conditions
used.  The doping concentration is estimated from reference samples to
be $n\approx 10^{16}$--$10^{17}$ cm$^{-3}$. To control the number of
electrons in the QDs, we rely on the photo-depletion (or negative
photoconductivity \cite{13a}) effect. Here, an electron-hole pair
photo-created in the AlGaAs barrier separates due to the space charge
field around the charged QD. The hole is attracted by the charged QD
and subsequently recombines with a QD-electron by emitting a photon,
while the electron neutralizes one of the ionized donors in the AlGaAs
barrier (Fig.  1b). Thus, photo-depletion removes electrons from the QD
at a rate $r_{\rm depl}\sim P_{\rm exc}$ depending on the absorbed
laser power $P_{\rm exc}$. Second, in order to reestablish equilibrium,
the electron bound at donors {\em hop back}\/ into the QD with
thermally activated rates $r_{\rm hop}(i)$ depending strongly on the
distance of each donor $i$ to the QD (Fig. 1c) \cite{14}. Thus, for
each excitation power $P_{\rm exc}$, the competition between
photo-depletion $r_{\rm depl}$ and back-hopping $r_{\rm hop}(i)$ will
determine the stationary QD charge $n_{\rm stat}(P_{\rm exc})$.
However, since the back-hopping rates $r_{\rm hop}(i)$ depend
sensitively on the exact impurity configuration, $n_{\rm stat}(P_{\rm
exc})$ will be unique for each QD.  With increasing $P_{\rm exc}$ the
QD will be continuously depleted until finally all charges are removed
(Fig.1d).

Figure 2a shows a typical excitation-power-dependent evolution of the
ground state transition of a single QD. In this evolution, two
complementary regimes are observed. The low power regime (LPR, $P_{\rm
exc}\le 2.5$ nW) is characterized by a global blue-shift of the QD
transitions, and the high power regime (HPR, $P_{\rm exc}>2.5$ nW) by a
red-shift. An upper estimate of the average number of excitons in the
QD as function of excitation power is given by $N_X\approx 0.06\;P_{\rm
exc}$ excitons/nW \cite{13a}. In the LPR, $N_X$ is very small
($0.0017\le N_{X,\rm LPR}\le 0.34$) and the integrated QD PL intensity
$I_{\rm QD}$ increases almost linearly with the excitation power,
$I_{\rm QD}\propto P_{\rm exc}$ (Fig.  2b). We thus conclude that the
strong spectral evolution in the LPR can not be due to multi-excitonic
effects. We tentatively ascribe the level shifts in the LPR to the
change in the number of surplus electrons. In what follows, this
interpretation will be confirmed by our detailed few-particle
calculations, which show that the optical spectrum must change whenever
an additional carrier is added to the dot because of the resulting
additional Coulomb interactions. However, in the HPR multi-exciton
effects are believed to govern the spectral evolution since both $N_X$
is rather high ($0.34<N_{X,\rm HPR}\le 33$) and $I_{\rm QD}$ saturates
in this regime (Fig. 2b).

The following observations show that the LPR is indeed governed by the
photo- depletion/back-hopping mechanism. First, using a red Ti-Sapphire
laser, we could selectively excite only the GaAs material in our
samples, virtually turning off the mechanism of photo-depletion.  In
such power-dependent measurements, the pronounced spectral evolution in
the LPR is absent (not shown here, see Ref. \cite{17}). Thus,
photo-depletion is essential for the occurrence of this spectral
evolution. Second, we performed two-color time-resolved pump and probe
PL measurements revealing relaxation times in the ms-range for the
spectral changes observed in the LPR \cite{18}. This is a clear
indication for the importance of the slow back-hopping process.
Finally, we mention the influence of temperature on the spectral
evolution in the LPR. Figure 3a shows the 10K-spectral evolution
obtained from another single QD of the same sample. Here, the
blue-shift observed in the LPR is much less pronounced than for the QD
studied in Fig. 2a, in spite of the larger range of excitation powers
involved. Within the photo-depletion/back-hopping model, this indicates
a smaller change in $n_{\rm stat}(P_{\rm exc})$ and an almost complete
depletion already at low $P_{\rm exc}$. We therefore assume that for
this specific dot the QD-donor distances are larger, leading to smaller
back-hopping rates $r_{\rm hop}(i)$. Thus, increasing the temperature
should enhance $r_{\rm hop}(i)$, giving rise to larger $n_{\rm
stat}(P_{\rm exc})$ at low $P_{\rm exc}$ and consequently to a more
ponounced blue-shift with increasing $P_{\rm exc}$. This is exactly
what we observe in the 40K-spectra of Fig.~3b.  More quantitatively, we
find similar spectra for $P_{\rm exc}=3$ pW at 10K and $P_{\rm exc}=7$
nW at 40K indicating a strong ($\sim$1000-fold) enhancement of the
$r_{\rm hop}(i)$ with increasing temperature. To summarize this point,
the dependence on excitation wavelength, the temporal evolution and the
influence of temperature all indicate that the model of
photo-depletion/back-hopping is appropriate for describing the observed
LPR.

Next, we address the question whether the fine structure in the LPR is
governed by {\em intrinsic}\/ QD multi-particle interactions or by {\em
extrinsic}\/ QD-impurity interactions.  Such extrinsic effects could be
Stark shifts or modifications of the optical selection rules due to
deformations of the QD-confined wavefunctions by electric fields of
ionized impurities close to the QD. Since intrinsic factors depend only
on the QD-potential, they should be reproducible from QD to QD.
Extrinsic factors, on the other hand, depend on the exact impurity
configuration and are therefore not reproducible.  Thus, we have
compared the evolution of the LPR-fine-structure of many QDs grown on
the same patterned substrate, and present here as an example those of
Figs. 2 and 3a,b. All QDs exhibit indeed remarkably similar evolutions
in the LPR. We can identify in both samples of Figs. 2 and 3 a number
of characteristic peaks or peak multiplets appearing at lower energies
as the excitation power is reduced. The numbers and relative energetic
positions of the individual peaks with respect to the highest energy
transition $X$ are identical (within 0.5--1 meV) for the two QDs
presented in Figs. 2 and 3 as well as for the 10 other QDs which we
studied in detail. Hence, these peaks or peak-multiplets are intrinsic
to the QD and can therefore be attributed to specific multi-particle
configurations. With increasing excitation power, electrons are
sequentially removed from the QD via the photo-depletion/back-hopping
mechanism. We therefore attribute each peak or peak-multiplet to a
certain electron occupation of the QD.

To investigate the multi-particle states of single dots, we performed
detailed theoretical calculations. We start from the single-particle
states, which are derived by numerically solving the 3D single-particle
Schrdinger equation within the envelope- function and effective-mass
approximations. The QD confinement is assumed to have cylindrical
symmetry, with the shape of the confining potential obtained from the
experimental AFM data \cite{12}; for the conduction- and valence-band
offsets we use material parameters for GaAs/AlGaAs. Without further
adjustment of parameters, we obtain an energy level splitting of
approximately 30 meV for electrons and 10 meV for holes. Next, the
many-particle Hamiltonian (containing all possible electron- electron,
electron-hole, and hole-hole Coulomb matrix elements) is expanded
within the basis of the 10 energetically-lowest single-particle states
for electrons and holes, respectively. Keeping the $\sim$100 Slater
determinants of lowest single-particle energies for electrons and
holes, respectively, we obtain the many-particle states by direct
diagonalization of the Hamiltonian matrix. This approach is similar to
those presented in Refs. \cite{19,20}, and will be discussed in detail
elsewhere.

Figure 4 shows luminescence spectra as computed for different numbers
of electrons and holes (see expressions on the right-hand side of each
spectrum) \cite{20}. For a single electron-hole pair confined in the QD
($X$, Fig.  4(c)), the luminescence originates solely from the decay of
the ground-state exciton. Our calculations reveal an energy splitting
between the (occupied) exciton ground-state and the (unoccupied) first
excited exciton state of 43 meV; this value matches well the
experimentally observed features from higher shells, and further
supports our theoretical choice of the confining QD potential. In the
HPR, we observe the appearance of a biexcitonic line ($2X$, Fig. 4(b))
which is shifted by $\sim$2 meV to lower photon energies; a further
red-shift is observed for the decay of the charged biexciton
($3e$-$2h$, Fig. 4(a)). In the LPR (Figs.  4(d-h)), the QD is populated
by only one hole and multiple electrons. When increasing the number of
electrons, the main emission peaks in the optical spectra monotonically
red-shift as a consequence of the modified interactions \cite{21,22},
and additional peaks appear on their low-energy side.

We used the good agreement between the theoretical results (Fig.4) and
the experimental spectra to identify the transitions due to the
different multi-particle states (as marked in Figs. 2 and 3). For
example, in both the theoretical as well as in the experimental
spectra, the excitonic line $X$ is followed in the LPR by a strongly
shifted, single line ascribed to the charged exciton $2e-h$. At still
lower powers, a weaker shifted line $3e-h$ with a weak satellite appear
at lower energies, and so on. Charged exciton states with up to five
electrons can thus be identified in the optical spectra.  We attribute
the minor differences regarding the peak positions in the experimental
and theoretical spectra (Figs. 2,3 and 4) to the uncertainties
regarding the detailed shape of the QD confinement potential. Finally,
to clarify the role of electric-fields of ionized impurities on the
few-particle states we have additionally performed model calculations
for a dot potential modified by an impurity in the surrounding, and
have found only minor variations in the optical transition energies
(e.g., less than 1 meV for a dot-impurity distance of 10 nm)
\cite{23}.

In conclusion, we have presented an experimental and theoretical
investigation of modulation-doped single QDs. We have demonstrated that
the mechanism of photo- depletion/back-hopping can be efficiently
used to control the number of electrons in the QD. The evolution of
single-QD PL spectra as function of excitation power was shown to be
governed by intrinsic QD-multi particle effects. The comparison with
theoretical spectra allowed us to identify the different characteristic
transitions of each QD-multi-particle state. In particular, we have
observed the formation of charged QD exciton states obtained by
sequentially adding electrons to the QD.

This work was supported in part by the Fonds National Suisse de la
Recherche Scientifique within PNR36 {\em Nanosciences},\/ by INFM
(Italy) through PRA--SSQI, and by the EC under the TMR Network {\em
Ultrafast Quantum Optoelectronics}.\/ U.H.  acknowledges support by the
EC through a TMR Marie Curie Grant.

\begin{figure}
\caption{
(a) Scanning electron microscope image of pyramidal QD-structures after
the back-etching process. (b)--(d) Schematical illustrations of the
mechanisms of (b) photo-depletion, (c) back-hopping of electrons and
(d) total depletion of the QD at high excitation power levels.
}
\end{figure}

\begin{figure}
\caption{
(a) Excitation power dependent evolution of the PL spectrum of a single
QD. A number of characteristic transitions for different multi-particle
configurations in the QD are identified (see text). The intensities are
normalized for better visibility. In order to underline the
absence/presence of low energy satellite peaks for the $2e$-$h/3e$-$h$
transitions, respectively, the intensities of the spectra left of the
dashed, vertical line have been enhanced by a factor of ten. (b)
Integrated PL intensity of the different characteristic transitions.
}
\end{figure}

\begin{figure}
\caption{
(a) 10K excitation power dependent PL evolution of different single QD
than shown in Fig. 2. (b) Same QD as in (a) measured at 40K. (see also
caption of Fig.2).
}
\end{figure}

\begin{figure}
\caption{
(a)--(h) Calculated spectra for different multi-particle configurations
in the QD; photon energy zero is given by the groundstate exciton $X$.
In the calculation of the luminescence spectra we assume that before
photon emission the interacting electron-hole states are occupied
according to a thermal distribution at temperature $T=10$K, and we
introduce a small broadening of the emission peaks accounting for
interactions with the dot environment (e.g., phonons, additional
carriers in the surrounding).
}
\end{figure}

\end{multicols}


\begin{thebibliography}{10}
\bibitem{13a} We use $N_X\approx N_{\rm phot}\;R\;\alpha\;\tau_{\rm
QD}$, with $N_{\rm phot}$ the number of incident photons, $T=0.7$ the
transmittance of the AlGaAs surface, $\alpha=0.063$ the ratio of the
integrated PL intensities of QD to the total pyramidal structure, and
$\tau_{\rm QD}\approx 0.5$ ns the lifetime of QD excitons.
\bibitem{23} Such small shifts of the optical transition energies allow
us to explain the experimentally observed spectral narrowing of
emission peaks in Fig. 2: With increasing $P_{\rm exc}$ more impurities
in the vicinity of the QD remain neutralized or are screened by
additional photo-excited carriers in the barrier, leading to smaller
electric-field fluctuations at the QD.
\end{thebibliography}
\end{document}